\let\Xdocument\document
\documentclass{iau}
\let\document\Xdocument

\usepackage{amsmath}
\usepackage{graphicx}
\usepackage{multirow}

\begin{document}

\lefttitle{Michele Trabucchi}
\righttitle{Long-Period Variables as distance and age indicators in the era of \textit{Gaia} and LSST}

\jnlPage{1}{7}
\jnlDoiYr{2023}
\doival{10.1017/xxxxx}

\aopheadtitle{Proceedings IAU Symposium}
\editors{R. de Grijs,  P. A. Whitelock \&  M. Catelan, eds.}

\title{Long-Period Variables as distance and age indicators in the era of \textit{Gaia} and LSST}

\author{Michele Trabucchi$^{1,2}$}
\affiliation{
$^{1}$  Dipartimento di Fisica e Astronomia Galileo Galilei,
        Università degli studi di Padova,
        Vicolo dell'Osservatorio 3, I-35122 Padova, Italy
\\
$^{2}$  Department of Astronomy,
        University of Geneva,
        Ch. Pegasi 51, 1920 Versoix, Switzerland
}

\begin{abstract}
Long-period variables are bright, evolved red giant stars showing periodic photometric changes due to stellar pulsation. They follow one or more period-luminosity and period-age relations, which make them highly promising distance indicators and tracers of young and intermediate-age stellar populations. Such a potential is especially interesting in view of the massive amount of data delivered by modern large-scale variability surveys. Crucially, these applications require a clear theoretical understanding of pulsation physics in connection with stellar evolution. Here, I describe an ongoing effort from our collaboration dedicated to the modelling of stellar pulsation in evolved stars, and how this work is impacting our capability of investigating long-period variables and exploiting them for other astrophysical studies. Furthermore, I present our ongoing work aimed at assessing the potential of semi-regular variables, an often neglected sub-type of long-period variables, to be distance indicators complementary to their better-known, more evolved counterparts, the Mira variables.
\end{abstract}


\maketitle

\section{Introduction}
\label{sec:Introduction}

Low- and intermediate-mass stars ($0.8\lesssim M/{\rm M}_{\odot}\lesssim 8$) approach the end of their lives through the Asymptotic Giant Branch (AGB) evolutionary phase, during which their envelope becomes unstable to stellar pulsation in low-order oscillation modes. This process is observed in the form of periodic stellar variability characterized by photometric amplitudes up to several magnitudes in visual bands and by periods of order of months up to a few years, hence the name Long-Period Variables (LPVs). This broad class of variable stars does not have a strict definition, and often extends to more massive or less evolved objects, such as core-He-burning red supergiants (RSGs) or stars approaching the tip of the red giant branch (RGB).

Whether or not an object is considered a LPV often depends on the purpose of an observing program, or on its instrumental capabilities. In particular, LPV catalogs from large surveys often include all giant stars with red colors that display photometric changes above the instrumental variability detection threshold. In the 1990s, this kind of surveys have revolutionized our understanding of LPVs (see, e.g., the review contribution by Igor Soszyński in these proceedings), and have revealed a striking degree of complexity in the behaviour of these stars that display multi-periodic variability caused both by pulsation in multiple distinct modes (possibly simultaneously) and by other processes, typically associated with the presence of a binary companion (see Micha{\l} Pawlak's contribution in these proceedings).

As many other pulsating stars, LPVs follow period-luminosity relations (see Patryk Iwanek's contribution in these proceedings) that make them excellent candidates to be distance indicators (see Miora Andriantsaralaza's contribution in these proceedings). The Mira variables are the sub-type of LPVs that has enjoyed the most attention in this sense: they are intrinsically very bright, especially at infrared (IR) wavelengths, and show rather regular variations with large photometric amplitudes, so that they are easily observable (even at relatively large distances) and identifiable (see Caroline Huang's contribution in these proceedings).

They also obey a long-known period-age relation by which they can be exploited as stellar chronometers and tracers of young stellar populations. Yet, these applications can in principle be extended to the smaller-amplitude progenitors of Miras, a sub-type of LPVs known as Semi-Regular Variables (SRVs). Often neglected because of their less-regular pulsation behaviour, which is challenging both on observational and theoretical grounds, they carry a wealth of additional physical information that is well within reach of modern time-domain astrophysical surveys, and is ready to be exploited.

More generally, the variability features (in particular the period) of LPVs are tightly connected with their physical properties, and provide a means for probing their structure and inferring their global stellar parameters. Therefore they are valuable tools for advancing our knowledge of the evolution of stars and stellar populations. Understanding the complexity of LPVs and characterizing the connection between their variability and their underlying properties is among the modern challenges for stellar modelling research, which has called for an effort to provide the solid theoretical grounds necessary for supporting the full exploitation of LPVs as tools for the exploration of our cosmic neighbourhood.

\section{Stellar pulsation models of LPVs: connecting theory with observations}
\label{sec:Stellar_pulsation_models_of_LPVs_connecting_theory_with_observations}
Stellar pulsation carries valuable information for constraining stellar structure and evolution models, provided that detailed knowledge is available about how the stellar structure determines the pulsation periods (as well as other variability-related observables) and how the pulsation behaviour is affected by evolutionary changes. For this reason, the modelling of stellar pulsation has long been an active field of research.

With a few exceptions \citep[e.g.][]{xiong_etal_2018} the theoretical investigation of pulsation in AGB stars has been rather dormant in the past few decades, whereas more recently it has seen an enormous progress owing to the development of bleeding-edge three-dimensional radiation-hydrodynamic codes that provide the most realistic picture of convection and its interaction with pulsation \citep[][; see also the Arief Ahmad's contribution in these proceedings]{ahmad_etal_2023}. An inherent shortcoming of these models stems from their extremely time-consuming nature, that makes them currently unfeasible for the exploration of a vast space of stellar parameters necessary for characterizing the AGB phase.

Instead, such an exploration is conveniently performed by adopting linear stellar pulsation codes, that sacrifice a detailed description of physical processes such as the pulsation-convection interaction in favor of computational efficiency. They can be used to construct grids of models to establish a connection between global stellar parameters and pulsation observables, which in turn can be used to link evolutionary calculations with variability in interpolation in the grid nodes. As an alternative, the relation between stellar structure and variability derived from such grids can be is conveniently encapsulated into easy-to-use analytic prescriptions, typically in the form of period-mass-radius relations (as the period is the main variability-related observable and mass and radius are the parameters affecting it the most).

The incomplete coverage of the space of stellar parameters, especially in terms of chemistry, has been the main shortcoming of the various AGB pulsation prescriptions available in literature since the widespread appearance of pulsation codes. For instance, \citet{marigo_girardi_2007} included variability information into AGB stellar evolutionary tracks by combining several distinct prescriptions, and resorting to a number of approximations to account for a space parameter coverage that was limited in mass, poor in metallicity, and null in terms of the carbon-to-oxygen number abundance ratio (${\rm C/O}$), the very parameter that characterizes the dramatic chemical evolution of these stars.

\subsection{A new grid of LPV pulsation models}
Motivated by the need of incorporating a description of long-period variability into evolutionary models of the Thermally-Pulsing AGB phase, with the aim of attaining a sound physical calibration of the latter \citep{marigo_etal_2017,pastorelli_etal_2019,pastorelli_etal_2020}, our group devoted an effort to the production of a new grid of AGB pulsation models with a wide coverage of the space of stellar parameters, including chemistry \citep{trabucchi_etal_2019}. 

This was achieved with the linear, radial, non-adiabatic, one-dimensional pulsation code described in \citet[][and references therein]{wood_olivier_2014}, upgraded to include up-to-date low-temperature gas opacities, with the inclusion of molecular contributions. Special care was paid to ensure the full consistency between the metal mixture adopted for the chemically-homogeneous envelope and that employed to compute the opacity tables (in contrast with a number of previous studies that simply assumed a scaled-solar metal mixture).

The final grid densely covers the mass range $0.6\leq M/{\rm M}_{\odot}\leq 7.0$, the luminosity range $2.5\leq\log(L/{\rm L}_{\odot})\leq5.0$, the range of initial metallicity (i.e. at the beginning of the AGB) $0.001\leq Z_{\rm init.}\leq0.017$, ${\rm C/O}$ between 0.3 (suitable for Hot-Bottom-Burning stars) to 5.0 (for C-rich stars), and various hydrogen abundances ($0.6\leq X\leq0.8$). Moreover, each model was computed for two different values of the core mass and three distinct values of the mixing length parameter (thereby offering appropriate coverage in terms of effective temperature). For each combination of such parameters the grid provides the period and growth rate (i.e. the degree of excitation) for five distinct pulsation modes, from the fundamental mode (FM) to the fourth overtone mode (4OM).

\begin{figure}
\centering
\includegraphics[width=0.75\textwidth]{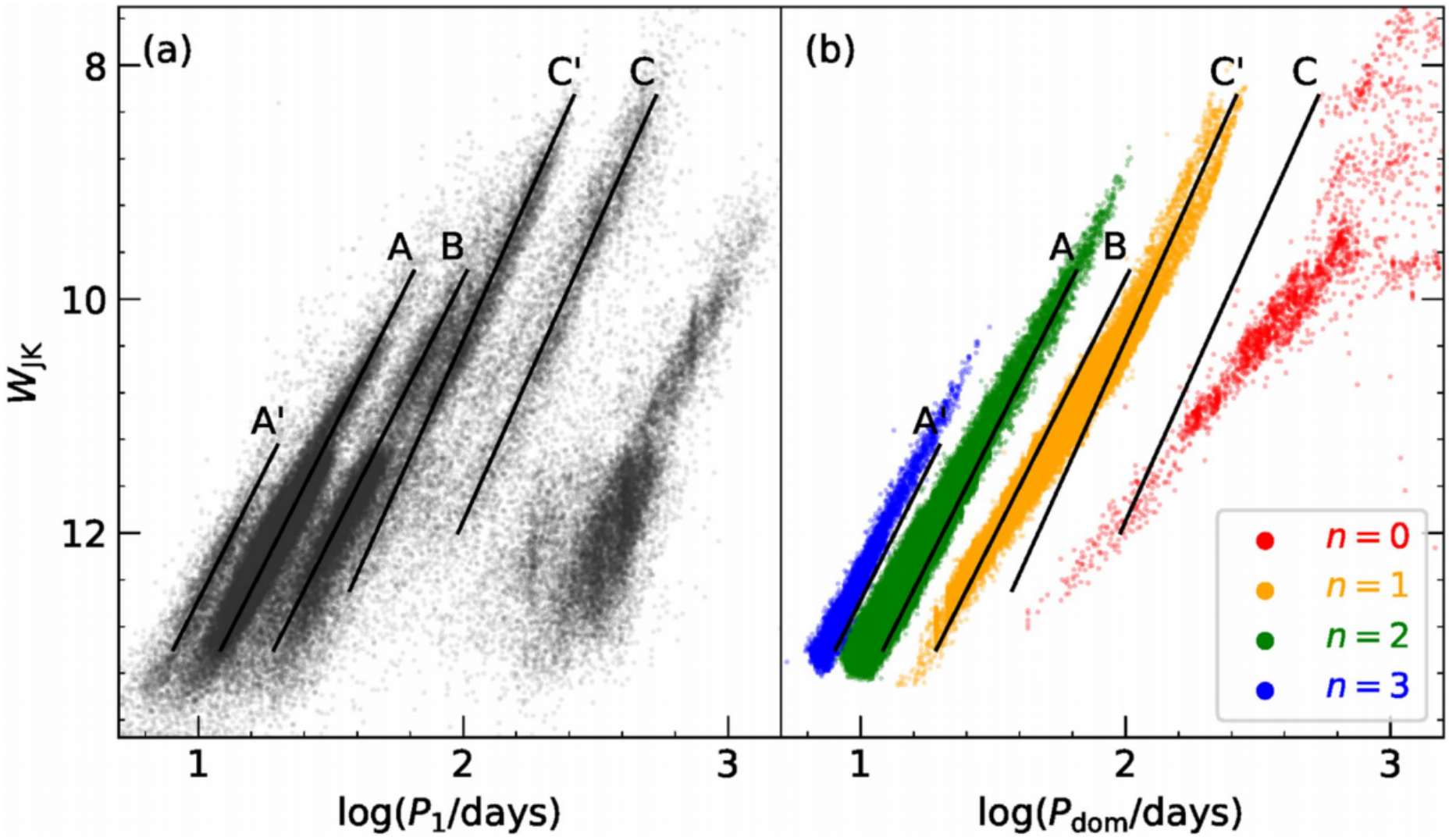}
\caption{Observed (left, using OGLE-III periods and 2MASS NIR magnitudes) and simulated (right) period-luminosity diagram of LPVs in the LMC. Lines and labels indicate the approximate location of the period-luminosity sequences (using the nomenclature of \citet{wood_2015}). Theoretical dominant periods (color-coded by pulsation modes) are obtained by combining a synthetic stellar population model with linear pulsation models of red giant envelopes. Adapted from \citet{trabucchi_etal_2017}.}
\label{fig:1}
\end{figure}

As a benchmark for checking the inter-operability between evolutionary tracks and pulsation models, as well as for testing the accuracy of the latter, we combined the pulsation grid with a synthetic model of the population of evolved stars in the Large Magellanic Cloud (LMC), obtaining a simulation of the LPV content of the LMC \citep{trabucchi_etal_2017}. A direct comparison between the near-IR period-luminosity diagram (PLD, see Fig.~\ref{fig:1}) of the LPVs observed in the LMC by the OGLE-III program \citep{soszynski_etal_2009} and its simulated counterpart confirmed the accuracy of the models, and led to a novel interpretation of the origin of the period-luminosity relations of LPVs. Indeed, the five\footnote{
    We consider here the main period-luminosity sequences, labelled A, B, C, C$^{\prime}$, and D according to the nomenclature derived from \citet{wood_2000}. However, it is worth recalling that the PLD has a rich fine structure, that various sequence split into sub-ridges, and that additional sequence appear depending on the adopted photometric system \citep[see][for a summary]{wood_2015}.
} period-luminosity sequences appearing in the observed PLD have been traditionally attributed to five distinct pulsation modes, although there has been disagreement on the exact modal assignment. Surprisingly, our simulated PLD displayed only four such sequences, with one of them (corresponding to 1OM periods) seamlessly covering the area occupied by the observed sequences B and C$^{\prime}$ combined.

The reason why the observed distribution of 1OM periods appears split into two sequences is caused by a selection bias linked with the way the PLD is customarily plotted. Indeed, only one period per star is shown in the PLD even in the case of multi-periodic variables such as the LPVs. This approach relies on the reasonable assumption that the `primary' period of each star (the one with the largest photometric amplitude in the observed light curve, or the strongest signature in the periodogram) is representative of the overall pulsation behaviour of that star. In turn, this relies on the implicit assumption that all observed periods are caused by pulsation. This is actually not the case for many LPV catalogs, among which is OGLE-III, that include Long Secondary Periods (LSPs). This kind of variability is caused by a cloud of dust, dragged by a sub-stellar companion, that periodically eclipses the red giant \citep{soszynski_etal_2021}, and can lead to large enough photometric variations to be identified as primary period. When this happens, the LSPs are drawn on the PLD leading to the appearance of the period-luminosity sequence C, whereas the corresponding pulsation periods are effectively ignored. Hence, this has the subtle side effect of systematically removing information from the PLD. More precisely, pulsation periods are selectively removed from the area of the PLD in the middle of the 1OM period distribution, creating an artificial gap along it and giving rise to the two distinct period-luminosity relations B and C$^{\prime}$.

This explanation brings into alignment the two main interpretations, in disagreement with each other, that have emerged since the discovery of the multiple period-luminosity relations of LPVs, and has been confirmed by asteroseismic studies \citep{yu_etal_2020}. Moreover, it opes the way for a better understanding of the nature of LSP-like variability and the relation between the period-luminosity relations and mass-loss on the AGB \citep{mcdonald_trabucchi_2019}.

These results are instructive in two ways. On the one hand, they clearly show the degree of complexity of the PLD of LPVs, resulting from their multiperiodic variability caused by the concurrence of multi-mode pulsation and binary-induced variations. On the other hand, it demonstrate the power of the approach based on the combination of evolutionary and/or synthetic population models with a suitably wide grid of pulsation models in order to improve our understanding of the period-luminosity relations of LPVs.

\subsection{The fundamental mode and the non-linear regime}
Another result emerging from the study by \citet{trabucchi_etal_2017} is the clear indication that linear pulsation models are not suitable for describing LPVs pulsating predominantly in the FM, such as Mira variables. Indeed, the linear approximation breaks down for these large-amplitude pulsators, a more accurate description of which requires non-linear hydrodynamic calculations. To address this issue, we used the hydrodynamic code described in \citet{keller_wood_2006} \citep[based on the work by][]{wood_1974} after applying the same opacity-related upgrades as the linear pulsation code, and performed the first systematic theoretical analysis of the behaviour of LPVs in the non-linear regime of pulsation \citep{trabucchi_etal_2021_nlp}.

The results from hydrodynamic calculations show that the period does not increase indefinitely with radius, as predicted by linear models. Instead, the sensitivity of the period to changes in radius decreases as the star expands, until eventually the period becomes independent of the radius. This confirmed previous indications \citep{yaari_tuchman_1996,lebzelter_wood_2005,kamath_etal_2010} that large-amplitude pulsation affects the structure of the pulsating layers of the envelopes, causing them to readjust to a condition of hydrodynamic equilibrium in which their (time-averaged) density is slightly larger than that of an identical model in hydrostatic equilibrium (i.e., static or pulsating with a small amplitude). In turn, the local increase in density causes the pulsation period to decrease (as can be understood by a naive application of the period-mean density relation), hence demonstrating the strongly non-linear nature of FM pulsation in LPVs.

\begin{figure}
\centering
\includegraphics[width=\textwidth]{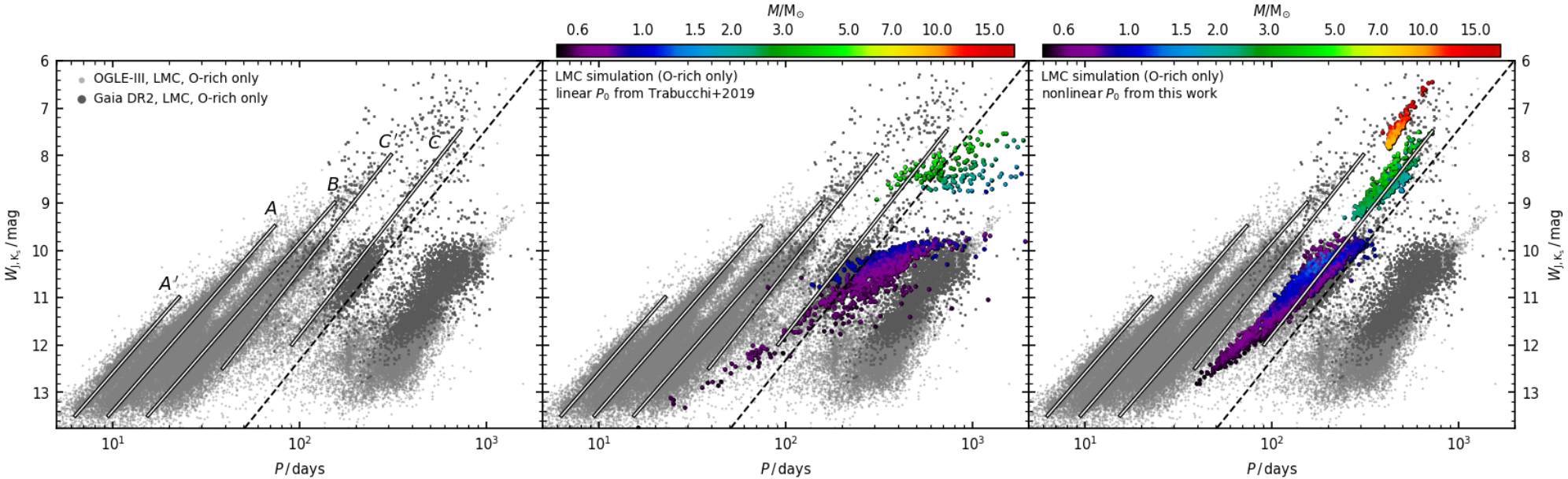}
\caption{Similar to Fig.~\ref{fig:2}, but limited to O-rich stars, and comparing observations from OGLE-III and \textit{Gaia} (left-most panel and gray dots in all panels) with inadequate FM periods predicted from linear pulsation models (central panel), and accurate FM periods from 1D hydrodynamic simulations (right-most panel). Simulated data points are color coded by the current stellar mass. Adapted from \citet{trabucchi_etal_2021_nlp}.}
\label{fig:2}
\end{figure}

It is worth pointing out that the non-linear period-mass-radius relation provided in \citet{trabucchi_etal_2021_nlp} is significantly more accurate than any such prescription based on linear pulsation models \citep[see, e.g.,][]{trabucchi_etal_2020}, as demonstrated by the improved agreement with observational data (see Fig.~\ref{fig:2}). This can be roughly quantified by considering one of the possible applications of such relations, that is the estimate of the stellar mass based on a knowledge of the FM period and of the stellar radius. By comparing Eq.~1 of \citet{trabucchi_etal_2021_nlp} with Eq.~12 of \citet{marigo_girardi_2007} one finds that the latter would cause the mass to be underestimated by 50-to-100\% in the case of a massive AGB star, whereas it would be overestimated by a factor 2-to-4 in the case of a short-period ($\sim 100$ days), relatively low-mass AGB star, and by twice that much for a similar star with a long period ($\sim 500$ days).

As the hydrodynamic code is computationally more demanding than its linear counterpart, the non-linear grid was limited to a single chemical composition ($Z=0.006$, O-rich). An extension to cover at least the same nodes of the linear grid is in progress as part of a work preparatory for the exploitation of LPV data from the LSST survey of the upcoming Vera Rubin Observatory \citep[see][]{daltio_etal_2022}. The results of the grid are planned to be publicly available and to include several products beyond the pulsation periods, such as template light curves and physical parameters (surface luminosity, radius and temperature) as a function of the phase throughout the pulsation cycle.

\section{The period-age relation of LPVs}
\label{sec:The_period_age_relation_of_LPVs}
LPVs have long been known to obey a period-age relation. Between the 1920s and the 1940s it was realized that Galactic Miras with relatively short periods display hotter kinematics compared with Miras with longer periods, and therefore must belong to comparatively older stellar populations \citep{merrill_1923,wilson_merrill_1942,feast_1963}. A crude but convincing theoretical explanation of this anti-correlation between period and stellar age stars from the period-luminosity relation. As longer-period Miras are intrinsically brighter than their shorter-period counterparts, they must be more massive, and hence they must be younger (at least on the average). Whereas in principle valid for all pulsation modes usually active in LPVs, the period-age relation is most useful in the case of the FM. Clearly, it does not only concern the Mira variables, but all LPVs pulsating in the FM, including many SRVs.

Throughout the decades, there have been multiple spikes of interest for the promising applications of the period-age relation, including the recent appearance of several studies focused on mapping the central regions of the Milky Way galaxy, a task for which Mira variables are especially suited (see e.g. the contribution by Megan Lewis in these proceedings). In contrast, there have been comparatively few theoretical studies of the period-age relation \citep[e.g.,][]{wyatt_cahn_1983,feast_whitelock_1987,eggen_1998}. Undoubtedly, the lack of an accurate way of computing FM periods as a function of global stellar parameters must have contributed to hampering such studies. This situation has changed owing to the non-linear prescription derived by Trabucchi et al. 2021a, that we adopted in the attempt of deriving a model-based period-age relation for FM-pulsating LPVs \citep{trabucchi_mowlavi_2022}.

To do so, we combined the results from hydrodynamic pulsation models with state-of-the-art stellar isochrones in order to simulate the distribution of these age-tracing stars in the period-age plane. In order to compare the result with observations, we took inspiration from the approach of \citet{grady_etal_2019}, who calibrated the period-age relation on Miras whose age is known from their membership in star clusters. In addition to the clusters used by \citet{grady_etal_2019} we also adopted a few Magellanic Clouds (MCs) clusters containing a large number of LPVs with measured periods (both Miras and SRVs), and carefully identified the most-likely pulsation modes associated with each period (retaining only FM modes) and the chemical type of each LPV.

\begin{figure}
\centering
\includegraphics[width=0.75\textwidth]{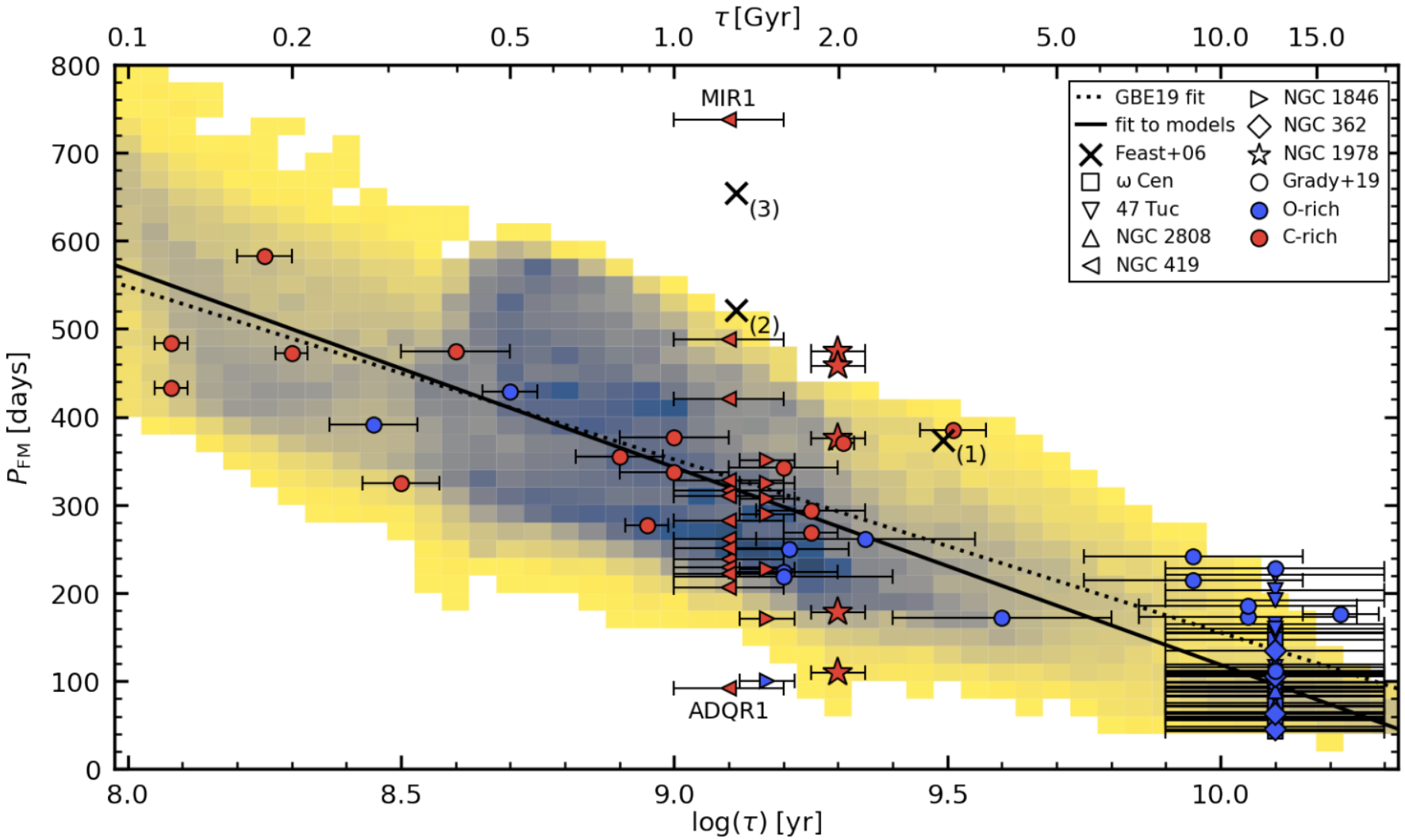}
\caption{Theoretical FM period-age relation (density map in the background and solid best-fit line) compared with observed periods of LPVs whose age is known from their membership in clusters. Stars belonging to different clusters are drawn with different symbol shapes, whereas the colors indicate different chemical type (blue for O-rich, red for C-rich). The observational best-fit by \citet{grady_etal_2019} is represented by the dashed line. Adapted from \citet{trabucchi_mowlavi_2022}.}
\label{fig:3}
\end{figure}

Models are in good agreement with observations (Fig.~\ref{fig:3}), both in terms of the best-fit period-age relation \citep[fully compatible with one derived by][]{grady_etal_2019} and of the spread of the relation, which turned out to be very large (for instance there is a $\sim$3 Gyr range for the age of a $\sim350$ days-period Mira). The spread in period at a given age (and vice versa) originates from the fact that the AGB phase, despite being almost instantaneous compared with previous evolutionary stages, is characterized by important structural changes, most importantly a substantial expansion that leads to large changes in periods. We note that the expansion of the envelope experiences an acceleration during the TP-AGB, whereas the period becomes progressively less sensitive to changes in radius as the latter increases. As a result, the theoretical period distribution at a given age is asymmetric, with a peak near its short-period end, in agreement with observations.

Such a spread in the period-age relation is far from ideal for using LPVs as age tracers unless properly characterized. In this respect, we note that advocates of this application traditionally consider only Miras as age indicators. Indeed, the Mira stage is so short compared to the duration of the AGB that it can effectively be treated as its endpoint, thereby selecting a much narrower range of periods at any given age. However, this approach has the downside of excluding a substantial number of other potential age tracers, namely the SRVs that pulsate in the FM and that are often very similar to Miras (see Sect.~\ref{sec:SRVs_as_distance_indicators}). Since the distinction between the two LPV sub-types is based only on their photometric amplitude, it is hard to apply to LPV models. Indeed, it requires them to be combined with a detailed (time-consuming) treatment of dust formation and radiation-hydrodynamics in the stellar atmosphere \citep[e.g.,][]{bladh_etal_2019}, whereas only an approximated description is currently used in the model.

It is also important to keep in mind that \citet{trabucchi_mowlavi_2022} examined the period-age relation using pulsation results that, strictly speaking, only apply to O-rich chemistry and LMC-like metallicity, while there is some indication \citep{trabucchi_etal_2019,trabucchi_etal_2021_nlp} that the chemical composition affects the instability of pulsation, i.e. the instability strip of the various modes, and thus the edges of the period-luminosity and period-age relations. The photometric amplitude of variability is also expected to be metallicity-dependent (especially at optical wavelengths) as it is determined by molecular absorption bands \citep{reid_goldston_2002} whose formation is favoured at higher metallicity. This poorly understood dependence might introduce a bias in the way Miras are distinguished from SRVs.

Our incomplete understanding of how the period-luminosity and period-age relations are affected by metallicity, and more in general by specific properties of any given stellar population (such as its star-formation history) is to be kept in mind when LPVs are adopted to map the Galaxy and its neighbourhood. An instructive example of this is offered by the results of \citet{sanders_etal_2022}, who used observations of Miras to characterize the properties of the Milky Way's nuclear stellar disc. The age distribution they derive is consistent with a number of previous studies but is not in agreement with the model-based period-age relation or with the relation derived by \citet{grady_etal_2019} from star clusters. Hopefully the extension of our grid of non-linear models to cover various chemical composition will help shedding light on these discrepancies, and enhance our capabilities of using LPVs as age indicators.

\section{SRVs as distance indicators}
\label{sec:SRVs_as_distance_indicators}

The large-scale optical microlensing surveys initiated at the end of last century (such as MACHO, EROS, MOA, OGLE) have been exceptionally beneficial to the study of variable stars. Among these surveys, the Optical Gravitational Lensing Experiment 
\citet{udalski_etal_1992}
has been especially valuable to understand LPVs. In particular, owing to its high cadence and photometric sensitivity, it allowed for the detailed characterization of the variability and the period-luminosity relations of SRVs and other low-amplitude LPVs. This is clearly illustrated by how OGLE enabled the identification of the OGLE Small-Amplitude Red Giants \citep[OSARGs, ][]{wray_etal_2004}, a sub-type of LPVs displaying a rich pattern of radial and non-radial oscillations, possibly bridging the gap between LPVs and solar-like oscillators on the RGB, that would otherwise be classified as irregular variables or even constant stars.

Current and upcoming time-domain astronomical surveys are expected to be comparable to OGLE for characterizing small-amplitude LPVs, whereas they can improve on OGLE in terms of sky coverage (e.g. \textit{Gaia}) or photometric depth (e.g. LSST). It is therefore interesting to assess whether SRVs\footnote{
    We adhere to the classification from the OGLE-III catalog, according to which the SRVs can approximately be identified to the non-Mira LPVs whose primary period lie on sequences C$^{\prime}$ or C.
} have the potential of being distance indicators complementary to Miras. This investigation is motivated by a number of arguments:
\begin{itemize}
    \item   they obey the same PL relation of Miras and/or an additional relation,
    \item   they are longer-lived than Miras, compared to which they are more numerous in any given stellar population (by roughly a factor 10 in the MCs) and trace older stellar populations (thereby representing a potential brighter alternative to RR Lyr variables),
    \item   they are less dusty than Miras, and thus less biased by dust absorption (in the optical) and emission (in the IR),
    \item   their photometric amplitudes are not too small to go undetected at optical wavelengths (where variability is observed), but small enough in the IR (where the PL relations are characterized) to reduce the scatter of the PL relation when it has to rely on single-epoch observations.
\end{itemize}

\begin{figure}
\centering
\includegraphics[width=\textwidth]{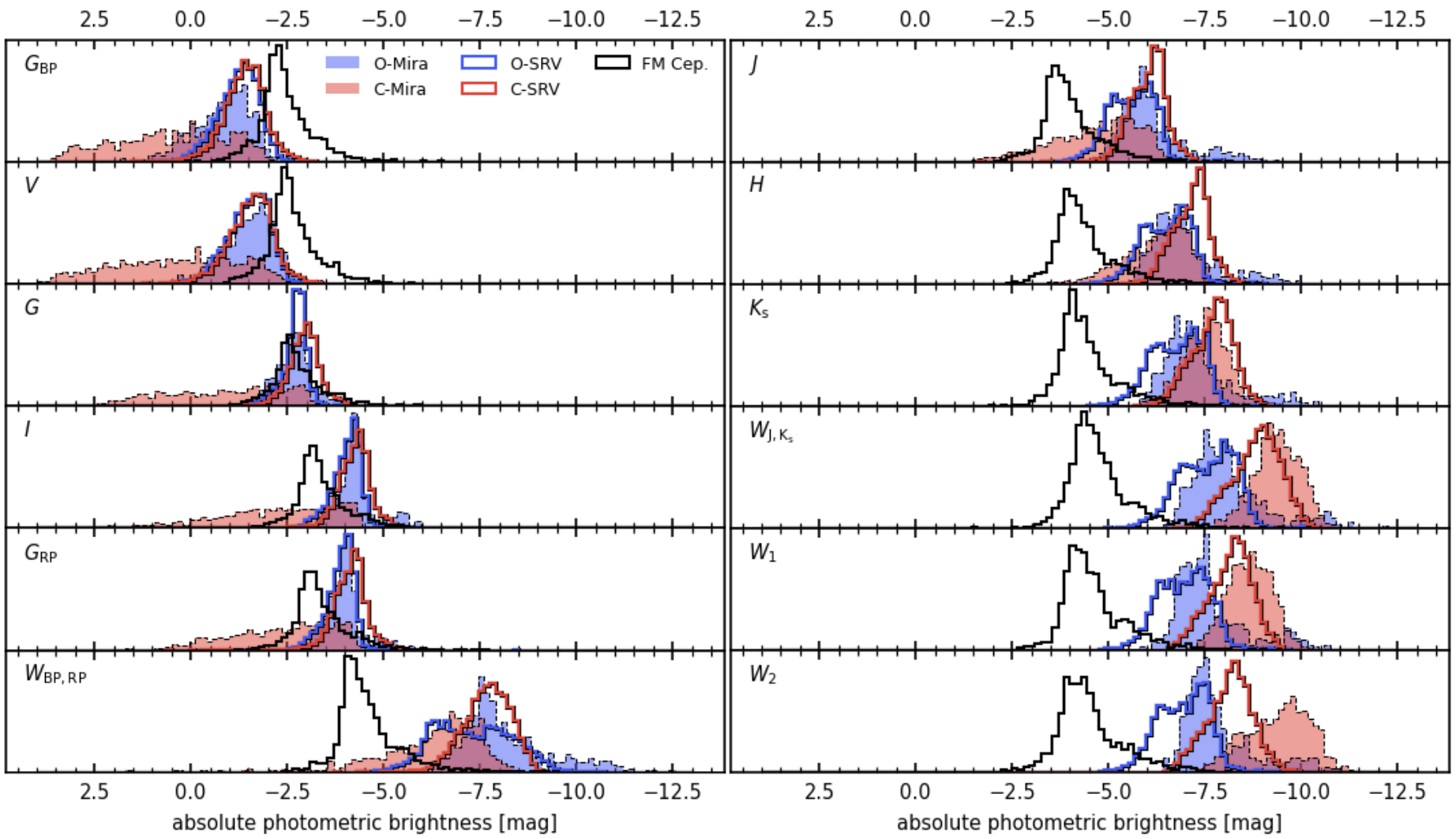}
\caption{Luminosity functions of LPVs and classical Cepheids (from OGLE-III and OGLE-IV, respectively) in different bands from blue to the mid-IR. A distinction is made between O-rich (blue) and C-rich (red) LPVs, as well as between Miras (filled histograms) and SRVs, while the solid black curves correspond to Cepheids. Each panel correspond to a distinct photometric filter: $G$, $G_{\rm BP}$ and $G_{\rm RP}$ from \textit{Gaia}, $V$ and $I$ from OGLE, $J$, $H$ and $K_{\rm s}$ from 2MASS, $W_1$ and $W_2$ from AllWISE. The \textit{Gaia} and 2MASS Wesenheit indices, $W_{\rm BP,RP}$ and $W_{J,K_s}$ are defined as in \citet{lebzelter_etal_2018}. Panels are sorted approximately by increasing central wavelength of the filters.}
\label{fig:4}
\end{figure}

It is also instructive to note that, while SRVs are typically fainter than classical Cepheids in blue and visual filters, their optical brightness is comparable in the optical, and SRVs are brighter in the IR (Fig.~\ref{fig:4}). Moreover, SRVs can be more numerous than Cepheids (e.g. in the LMC). What makes SRVs less appealing than Miras as distance indicators, besides their smaller amplitude (that is becoming less of a problems owing to the improvements in observational instrumentation and data reduction pipelines) is their lesser degree of regularity associated with their multi-periodic nature. While this makes them more challenging to deal with, it should be noted that any period beyond the first carries additional physical information that is worth exploiting.

In view of such considerations, we undertook a series of exploratory studies aimed at assessing the potential of SRVs, starting with a characterization of these stars in relation with Miras \citep{trabucchi_etal_2021_srv} based on OGLE-III observations of LPVs in the MCs. This analysis resulted in the identification of four different sub-groups of SRVs, depending on whether they are mono- or bi-periodic, and whether they pulsate predominantly in the FM or 1OM. The mono-periodic FM-pulsating SRVs are especially similar to Miras, except they display a smaller photometric amplitude, confirming previous findings \citep[e.g.,][and references therein]{lebzelter_hinkle_2002}. In particular, a sub-set of these stars follow the same sequences as Miras in the PLD as well as in the period-amplitude diagram, without discontinuity. This suggests that SRVs pulsating only in the FM should be treated as the same type of object from a physical perspective, and possibly for the purpose of distance measurements.

A direct benefit of adopting this criterion when selecting distance indicators among LPVs is that the pool from which they are drawn is roughly twice as large. Clearly, this is a rough estimate that holds in a statistical sense and, strictly speaking, only for a population of LPVs very similar to that of the MCs, and for observations obtained with an instrumental setup very close to that of the OGLE program. Nonetheless, there is a net advantage in terms of number of sources that can be observed, and it can be enhanced if the criterion is made more inclusive by accepting also SRVs with smaller amplitudes, bi-periodic, or pulsating in the 1OM.

However, the expansion of the set of LPVs statistically available to be used as distance indicators is not necessarily beneficial in terms of the performance of the PLR as a standard candle. For the purpose of our study, that is a comparative analysis of a large number of PLRs defined in different ways, we quantify their performance in terms of their capacity of correctly predicting the brightness of the sample of LPVs from which they were derived. In other words, we define a sample of LPVs (for instance, a reference set including only Miras, or a set consisting of Miras and a certain group of SRVs, and so on), derive the best-fit PLR of the sample, and compute the brightness residuals, i.e. the differences between the true brightness and the one predicted by the best-fit PLR. We then characterize the distribution of the residuals in terms of its width (that is related to the precision of the PLR) and its shape (i.e. how symmetric and well-centered it is, that is related to the accuracy of the PLR).

In general, the performance of the PLR becomes degraded if the sample is blindly extended to include all SRVs of any given group. Indeed, the SRVs that pulsate in the FM suffer from a larger scatter than Miras around the PLR. However, a more selective addition of sources can is indeed beneficial. For instance, SRVs can be added to the sample by using a lower amplitude threshold than that traditionally used in the definition of Miras. There exists an optimal level of the threshold that maximizes both the precision and accuracy of the PLR. The optimal threshold depends on several factors, such as the photometric band the PLR is observed in, the SRV groups included in the sample, the chemical type of the adopted LPVs, and the pulsation mode (FM or 1OM). In the case of some photometric filters, this approach only leads to marginal improvements (or no improvement at all), and does not justify the use of the extended PLR, whereas the improvement is evident, for instance, in the NIR and in some MIR bands.

This analysis shows that the traditional separation between Miras and SRVs, based purely on amplitude, is not necessarily physically meaningful, nor it is optimal in terms of the performance of the PLR as a distance indicator. Instead, for the purpose of distance (and possibly age) determinations, one could consider a more fluid, efficiency-driven selection criterion. It is worth recalling that the amplitude-based definition is also prone to a number of biases. A relevant example, discussed by \citet{lebzelter_etal_2022}, concerns a comparison between sources in the \textit{Gaia} DR3 catalog of LPV candidates and their best matches in the ASAS-SN catalog \citep{jayasinghe_etal_2021}. There exists a discrepancy between the photometric amplitude published in the two catalogs, with the largest differences localized near the Galactic Bulge and plane. In this area, many sources with a relatively small $V$-band amplitude according to ASAS-SN are shown to have a large $G$-band amplitude by \textit{Gaia}. The reason behind these differences is that, owing to its higher spatial resolution, \textit{Gaia} light curves are less affected by the compression effect caused by crowding \citep[e.g.][]{riess_etal_2020}. As a result of this subtle effect, a number of Miras are incorrectly classified as SRVs in the ASAS-SN catalog.

\begin{figure}
\centering
\includegraphics[width=0.6\textwidth]{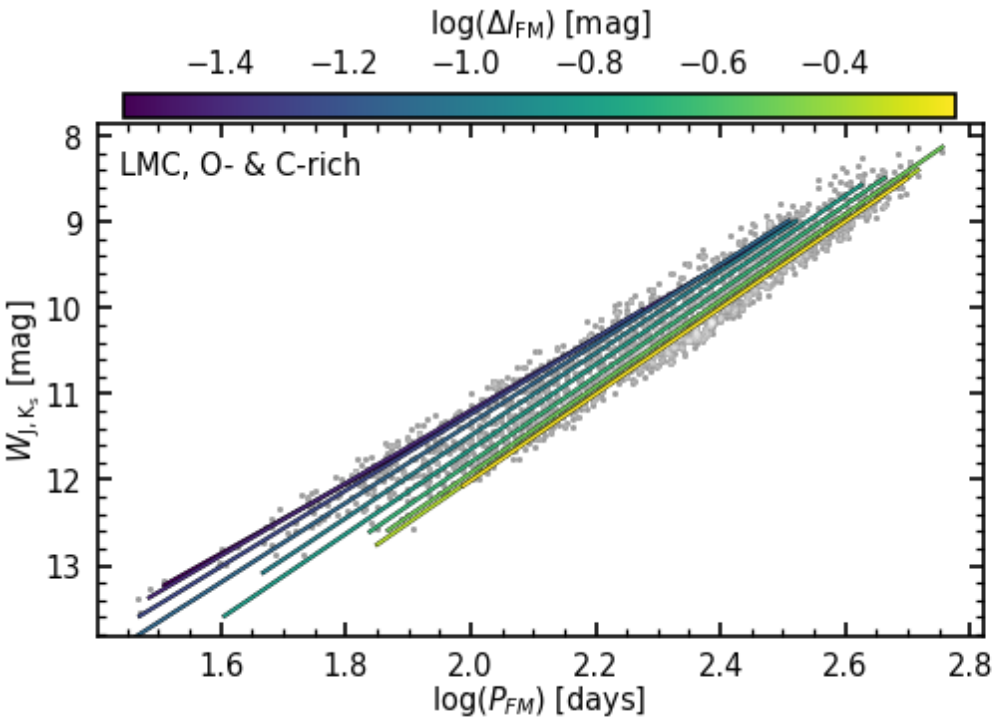}
\caption{Example best-fit analytic NIR period-luminosity relations for a sample of SRVs in the LMC in different amplitude bins. Each line represent a different relation, and is color-coded by the mean FM amplitude $\Delta I_{\rm FM}$ in the OGLE $I$ band of the corresponding bin. The relations, with varying slope and zero-point, span the full width of the observed period-luminosity sequence.}
\label{fig:5}
\end{figure}

At the same time, photometric amplitudes (as well as periods) are effectively independent of distance and extinction, and carry valuable information provided that effects of crowding are properly accounted for. One way of doing so is based on the observation that the amplitude of variability increases with period and luminosity, and in particular it increases \textit{across} the PLRs \citep[e.g. figure 1 of][]{soszynski_etal_2013}. Each PL sequence can thus be treated as a family of lines having slope and zero-point that vary with amplitude, thereby spanning the sequence's entire width (Fig.~\ref{fig:5}). In other words, the coefficients of the PLR can be thought as amplitude-dependent. This perspective turns the amplitude into a useful second parameter of the PLR, in a fashion that is similar to the introduction of a color term. A similar operation can be done by using the secondary period of bi-periodic SRVs (or the ratio of their period) as the second parameter.

By deriving the best-fit PLR in different amplitude bins for a given sample of LPVs, and then fitting the dependence of the PLR coefficients on the amplitude, we derived various expressions for the period-luminosity-amplitude relation (PALR) whose performance we can characterization of the PLR performance with the same framework described above. The results suggest that this approach effectively accounts for the scatter of the SRVs around the PLR, canceling (at least partially) its negative effect on the performance of the PLR. Compared to the PLR in its traditional form, the PALR displays a distribution of residuals that is both narrower and better-centered, indicating a higher degree of precision and accuracy. The performance is comparable to that obtained when using a PLR limited to Miras or to LPVs with amplitude above an optimized threshold, but can be defined to include a substantially larger set of sources.

\section{Conclusions}
\label{Conclusions}
Long-period variables are powerful astrophysical tools, whose applications range from inferring stellar parameters from variability observations, constraining stellar evolution models, tracing stellar populations and estimating astronomical distances. In order to fully exploit the potential of these stars, a solid theoretical understanding is needed, as well as pulsation models with accurate predictive power. While promising, highly-realistic 3D hydrodynamic simulations of evolved red giant envelopes are being successfully developed, the grid-based approach involving flexible 1D models has proven ideal to widely explore the space of stellar parameters and establish a sound connection between the pulsation of LPVs, their structure and their evolution. In particular, it allowed for a novel interpretation of their period-luminosity relations, showing a higher degree of complexity than expected from previous studies, and providing new insight on the link between pulsation, mass-loss and long secondary periods (whose nature is still debated).

Research in this direction also showed the inadequacy of widespread prescriptions, based on linear pulsation models, adopted to compute fundamental-mode periods of LPVs from their global stellar parameters. A relation based on the result of 1D hydrodynamic pulsation model is able to overcome these shortcomings, accurately describing the period of large-amplitude Mira variables. This new relation prompted an exploratory, theory-based study of the period-age relation of these stars, that is in agreement with observations of LPVs in clusters, but at odds with evidence from Bulge stars, pointing towards the need of more detailed studies of the connection between pulsation and chemical composition.

Finally, the improvement in the capacity of modelling the pulsation of bright red giants opens new avenues for using semi-regular variables, alongside Miras, as potential distance indicators. SRVs have long been neglected for this application, despite following the same period-luminosity relation of Miras (and other ones) because of their smaller amplitude and lesser regularity. Yet, they have a number of properties that motivate assessing their potential as standard candles, such as their being more numerous than Miras and present in older stellar populations. Preliminary results indicate that some SRVs that are especially similar to Miras can help improve their performance for distance determination, and that similar results can be obtained by including information on photometric amplitude of variability as a second parameter in the period-luminosity relation.

\vspace{1cm}

{\bf Question (Whitelock):} I wonder if you are going to be looking in detail at the somewhat more massive HBB LPVs, that fall above the standard fundamental PLR. These could be important as distance indicators in the JWST era.

{\bf Answer:} The current grid extends up to $7 M\odot$ and extrapolation of the non-linear period-mass-radius prescription shows good agreement with observations, which is encouraging. We aim to extend the grid further up in mass in the future version of the grid in order to study HBB and super-AGB stars.

{\bf Question (Skowron):} Have you tried calculating individual distances for SRVs with these relations? I'm curious what errors on individual distances we can expect and how useful these variables would be as, e.g., Galactic structure tracers.

{\bf Answer:} Currently our work is only exploratory and a differential/comparative analysis. We derive the period-luminosity relations of Miras, then repeat after including SRVs in the sample, we look at the differences and see if there are improvements. We have not applied this to measure distances yet.

{\bf Question (Eyer):} A question about the use of radial velocities time series of LPVs incoming from Gaia.

{\bf Answer:} RVs from Gaia will be very useful for a number of applications. The one I'm most interested in is the calibration of the tricky free parameters in the pulsation models. Achieving such a calibration will further improve the predictive power of the models.

\end{document}